\documentclass[10pt]{article}
\usepackage[utf8]{inputenc}
\usepackage{fullpage}
\usepackage{geometry}
\usepackage{amsthm,amsmath,amssymb}
\usepackage{xcolor}
\usepackage{tikz}
\usetikzlibrary{arrows.meta,decorations.markings}
\tikzset{midarrow/.style={postaction={decorate},decoration={markings,mark=at position 0.5 with {\arrow{Stealth[length=2mm]}}}}}
\definecolor{pone}{HTML}{534AB7}
\definecolor{ptwo}{HTML}{0F6E56}
\usepackage{natbib}
\usepackage[colorlinks=true, linkcolor=blue, urlcolor=blue, citecolor=red, pagebackref, bookmarksnumbered]{hyperref}

\theoremstyle{plain}
\newtheorem{theorem}{Theorem}
\newtheorem{lemma}[theorem]{Lemma}
\newtheorem{observation}[theorem]{Observation}

\theoremstyle{definition}
\newtheorem{definition}[theorem]{Definition}

\title{Directed disjoint shortest paths via guillotine cuts}
\author{William Lochet\thanks{LIRMM, Universit\'e de Montpellier, CNRS, Montpellier, France. \texttt{william.lochet@gmail.com}}
\and M. S. Ramanujan\thanks{Department of Computer Science, University of Warwick, Coventry, United Kingdom.\newline \texttt{R.Maadapuzhi-Sridharan@warwick.ac.uk}}
\and Saket Saurabh\thanks{The Institute of Mathematical Sciences, Chennai, India, and Department of Informatics, University of Bergen, Bergen, Norway. \texttt{saket@imsc.res.in}}}
\date{}
\newcommand{\cP}{\mathcal{P}}
\newcommand{\cO}{\mathcal{O}}
\begin{document}
\maketitle

\begin{abstract}
We present a simple algorithm that solves the directed disjoint shortest paths problem with $k$ terminal pairs in time $n^{\cO(k^2)}$ on $n$-vertex digraphs, and we sketch how to improve its running time to $n^{\cO(k\log k)}$.

\medskip
\noindent\textbf{AI Statement.} The proof of Theorem~\ref{thm:main} was discovered by ChatGPT Astra during conversations with the authors on related problems. Following suggestions from the authors, it was able to improve the running time to $n^{\cO(k\log k)}$. This note presents the proof in a simplified and self-contained form that can be taught even in undergraduate courses. Claude Opus was used during the writing. 
\end{abstract}

\section{Introduction}

Given a digraph $D$ and $k$ pairs of vertices $(s_1,t_1),\dots,(s_k,t_k)$, the \emph{$k$-directed disjoint shortest paths problem} ($k$-DDSPP) asks whether there exist pairwise vertex-disjoint directed paths $P_1,\dots,P_k$ such that, for every $i\in[k]$, $P_i$ is a shortest $s_i$--$t_i$ path in $D$.

This is a variant of the classical $k$-disjoint paths problem, where the paths need not be shortest paths. In undirected graphs, the $k$-disjoint paths problem is FPT parameterized by $k$ by the celebrated result of Robertson and Seymour~\citep{RobertsonS95}, whereas in directed graphs it is already NP-complete for $k=2$~\citep{FortuneHW80}.

The undirected version of the problem was introduced by Eilam-Tzoreff~\citep{EilamTzoreff98}, who showed that it is NP-complete when $k$ is part of the input, gave a polynomial-time algorithm for $k=2$, and conjectured that the problem is polynomial-time solvable for every fixed $k$. This conjecture was confirmed more than twenty years later by the first author~\citep{Lochet21}, who gave an algorithm running in time $n^{\cO(k^{5^k})}$.
The directed version was introduced by Bérczi and Kobayashi~\citep{BercziK17}, who gave a polynomial-time algorithm for $k=2$. They also observed that Schrijver's algorithm for the $k$-disjoint paths problem in planar digraphs~\citep{Schrijver94} can be adapted to the shortest-path version, giving a polynomial-time algorithm for every fixed $k$ when the input digraph is planar. In general digraphs, however, the complexity of the problem remained open already for $k=3$.

In this note, we give a surprisingly simple algorithm for $k$-DDSPP running in time $n^{\cO(k^2)}$.

\begin{theorem}\label{thm:main}
For every fixed $k$, $k$-DDSPP can be solved in time $n^{\cO(k^2)}$ on $n$-vertex digraphs.
\end{theorem}

In fact, as we will sketch in Section \ref{sec:klogk}, the proof can be improved to $n^{\cO(k\log k)}$. We believe that the simplicity of Theorem \ref{thm:main} is the main contribution of this paper and chose to focus on explaining it clearly. 
The algorithm also applies to undirected graphs which greatly improves the best known running time in that case. In both cases, it can be adapted to positive arc lengths, but we restrict ourselves to unit lengths for simplicity. 

\section{Distance embedding and guillotine cut}

Let $D$ be a digraph with vertex set $V$ and $n$ vertices, and let $(s_1,t_1),\dots,(s_k,t_k)$ be $k$ pairs of vertices of $D$. For $i\in[k]$ and $v\in V$, let $d_i(v)$ denote the distance from $s_i$ to $v$, with $d_i(v)=n$ if $v$ is not reachable from $s_i$. Let $D_i$ be the subgraph of $D$ formed by the arcs that lie on some shortest $s_i$--$t_i$ path. Classically, $D_i$ is acyclic, and an $s_i$--$t_i$ path of $D$ is shortest if and only if it is a path of $D_i$. We assume that the $2k$ terminals are pairwise distinct. A \emph{solution} is a tuple $\cP=(P_1,\dots,P_k)$ of pairwise vertex-disjoint paths such that $P_i$ is an $s_i$--$t_i$ path of $D_i$ for every $i\in[k]$. The arcs of $P_1,\dots,P_k$ are the \emph{solution arcs}.

\begin{observation}\label{obs:step}
Let $i\in[k]$ and let $uv$ be an arc of $D_i$. Then $d_i(v)=d_i(u)+1$, and $d_j(v)\le d_j(u)+1$ for every $j\in[k]$.
\end{observation}

\begin{proof}
The equality holds because $uv$ lies on a shortest path from $s_i$. The inequality is the triangle inequality if $u$ is reachable from $s_j$, and otherwise $d_j(u)=n\ge d_j(v)$.
\end{proof}

Note that $d_j(v)$ may be much smaller than $d_j(u)$: along an arc of $D_i$, the distance $d_j$ increases by at most one but can decrease arbitrarily.

Let us consider the embedding $\phi:V\to\{0,\dots,n\}^k$ defined by $\phi(v)=(d_1(v),\dots,d_k(v))$. Its main role is to cluster the vertices of $D$ into boxes, in order to design a guillotine cut approach. A \emph{box} is a product $B=I_1\times\dots\times I_k$, where each $I_i$ is a nonempty interval of integers contained in $\{0,\dots,n\}$, and $|I_i|$ is the \emph{side length} of $B$ in coordinate $i$. We write $|B|=\prod_{i=1}^k|I_i|$ for the number of points of $B$. Abusing notation, we say that a vertex $v$ is in $B$ if $\phi(v)\in B$; note that many vertices may have the same image. Every vertex is in the \emph{root box} $B_0=\{0,\dots,n\}^k$. An arc \emph{crosses} $B$ if exactly one of its endpoints is in $B$. It will be important to specify which face of the box is crossed; the following definition is combinatorial rather than geometric.

\begin{definition}
Let $B=I_1\times\dots\times I_k$ be a box, let $e$ be an arc crossing $B$, and let $z$ be the endpoint of $e$ that is not in $B$. For $i\in[k]$, the arc $e$ \emph{crosses the upper $i$-face} of $B$ if $d_i(z)>\max I_i$, and it \emph{crosses the lower $i$-face} of $B$ if $d_i(z)<\min I_i$.
\end{definition}

Since $z\notin B$, some coordinate $i$ satisfies $d_i(z)\notin I_i$. Hence every arc crossing $B$ crosses at least one of the $2k$ faces of $B$, possibly several. The next definition is the standard guillotine cut.

\begin{definition}
Let $B=I_1\times\dots\times I_k$ be a box, let $a\in[k]$, and let $c$ be an integer with $\min I_a\le c<\max I_a$. \emph{Cutting $B$ at $(a,c)$} gives the two boxes $B^-$ and $B^+$ obtained from $B$ by replacing $I_a$ with $\{x\in I_a : x\le c\}$ and with $\{x\in I_a : x\ge c+1\}$, respectively. An arc $uv$ \emph{spans} $(a,c)$ if $\min(d_a(u),d_a(v))\le c<\max(d_a(u),d_a(v))$.
\end{definition}

Every vertex of $B$ lies in exactly one of $B^-$ and $B^+$, and $|B^-|,|B^+|<|B|$.

Finally, given a solution $\cP$, a box $B$ is \emph{good} for $\cP$ if each of the $2k$ faces of $B$ is crossed by at most $5k$ solution arcs. Since every arc crossing $B$ crosses at least one face of $B$, at most $10k^2$ solution arcs cross a good box. The root box $B_0$ is good for every solution, since no arc crosses it. The main idea is to show that a good box can always be cut into two smaller good boxes (Lemma~\ref{lem:box_cut}). This enables a dynamic programming approach, in which a state consists of a box $B$ together with at most $10k^2$ arcs, the candidates for the solution arcs crossing $B$.

\begin{lemma}\label{lem:movement}
Let $P=P_j$ be a path of a solution, let $B=I_1\times\dots\times I_k$ be a box with largest side length $L$ containing a vertex of $P$, and let $Q$ be the subpath of $P$ from its first to its last vertex in $B$. Then for every $a\in[k]$,
\[
\sum_{uv\in A(Q)}|d_a(v)-d_a(u)|\le 3(L-1).
\]
\end{lemma}

\begin{proof}
Let $x$ and $y$ be the first and last vertices of $Q$. Since $d_j$ increases by exactly one along each arc of $Q$ and $d_j(x),d_j(y)\in I_j$, the path $Q$ has at most $|I_j|-1\le L-1$ arcs. Let $R$ be the sum of $d_a(v)-d_a(u)$ over the arcs $uv$ of $Q$ with $d_a(v)>d_a(u)$, and $F$ the sum of $d_a(u)-d_a(v)$ over the arcs $uv$ of $Q$ with $d_a(v)<d_a(u)$. In other words, $R$ and $F$ are the total increase and the total decrease of $d_a$ along $Q$, and the sum in the statement is $R+F$.

By Observation~\ref{obs:step}, every arc $uv$ of $Q$ satisfies $d_a(v)-d_a(u)\le 1$. Since $Q$ has at most $L-1$ arcs, this implies $R\le L-1$. Moreover, $R-F=d_a(y)-d_a(x)\ge-(L-1)$ because $d_a(x),d_a(y)\in I_a$. Hence $F\le R+L-1\le 2(L-1)$, and $R+F\le 3(L-1)$.
\end{proof}

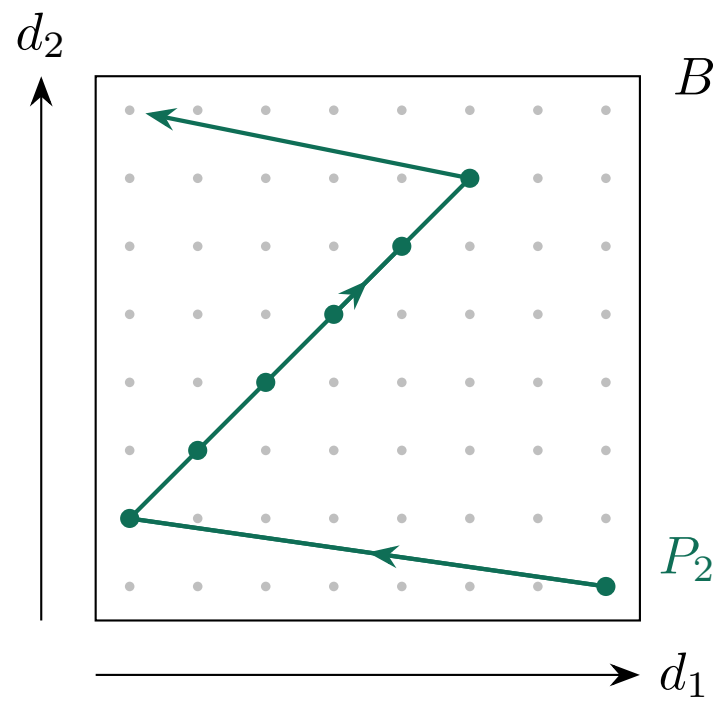
\begin{figure}[t]
\centering
\begin{tikzpicture}[scale=0.45, >={Stealth[length=2mm]}]
  \foreach \x in {0,...,7} \foreach \y in {0,...,7} \fill[black!25] (\x,\y) circle (2pt);
  \draw (-0.5,-0.5) rectangle (7.5,7.5);
  \node at (8.3,7.5) {$B$};
  \draw[->] (-0.5,-1.3) -- (7.5,-1.3) node[right] {$d_1$};
  \draw[->] (-1.3,-0.5) -- (-1.3,7.5) node[above] {$d_2$};
  \draw[ptwo, thick, ->, shorten >=3pt] (7,0) -- (0,1) -- (1,2) -- (2,3) -- (3,4) -- (4,5) -- (5,6) -- (0,7);
  \draw[ptwo, thick, midarrow] (7,0) -- (0,1);
  \draw[ptwo, thick, midarrow] (3,4) -- (4,5);
  \foreach \p in {(7,0),(0,1),(1,2),(2,3),(3,4),(4,5),(5,6)} \fill[ptwo] \p circle (4pt);
  \node[ptwo] at (8.2,0.4) {$P_2$};
\end{tikzpicture}
\caption{A path of $D_2$ in a box $B$, with each vertex $v$ drawn at $(d_1(v),d_2(v))$: along each arc it can move one to the right and arbitrarily far to the left.}
\label{fig:movement}
\end{figure}

In other words, between its first and last visit to $B$, the path $P$ moves by at most $3(L-1)$ along any distance $d_a$, however many times it leaves $B$ in between (Figure~\ref{fig:movement}).

\begin{lemma}\label{lem:box_cut}
Let $\cP$ be a solution and let $B$ be a box that is good for $\cP$ and has a side of length at least $2$. Then $B$ can be cut into two boxes $B^-$ and $B^+$ that are good for $\cP$.
\end{lemma}
 
\begin{proof}
Write $B=I_1\times\dots\times I_k$, let $a$ be a coordinate such that $L=|I_a|$ is the largest side length of $B$, and let $C=\{c\in\mathbb{Z} : \min I_a\le c<\max I_a\}$ be the set of values $c$ for which $B$ can be cut at $(a,c)$. Let $c\in C$, and let $B^-$ and $B^+$ be obtained by cutting $B$ at $(a,c)$. Every face of $B^-$ other than its upper $a$-face is defined by the same constraint as the corresponding face of $B$, so it is crossed by at most $5k$ solution arcs, since $B$ is good. Indeed, if a solution arc crosses such a face of $B^-$, its endpoint outside $B^-$ is also outside $B$, so the arc crosses the corresponding face of $B$. The same holds for every face of $B^+$ other than its lower $a$-face.

It remains to choose $c$ so that the upper $a$-face of $B^-$ and the lower $a$-face of $B^+$ are each crossed by at most $5k$ solution arcs. A solution arc crossing one of these two faces has an endpoint $w$ in $B$, and its other endpoint $z$ satisfies $d_a(w)\le c<d_a(z)$ or $d_a(z)\le c<d_a(w)$. Hence it has an endpoint in $B$ and spans $(a,c)$. Therefore it suffices to find $c\in C$ such that at most $5k$ solution arcs with an endpoint in $B$ span $(a,c)$.

Note that $|C|=L-1\ge 1$. An arc $uv$ spans $(a,c)$ for exactly $|d_a(u)-d_a(v)|$ integers $c$. Let $P$ be a path of $\cP$ with a vertex in $B$, and let $Q$ be as in Lemma~\ref{lem:movement}. The arcs of $P$ with an endpoint in $B$ are the arcs of $Q$, the arc entering $Q$ and the arc leaving $Q$. By Lemma~\ref{lem:movement}, the arcs of $Q$ span $(a,c)$ for at most $3(L-1)$ values of $c$ in total, and each of the two other arcs, whose length along $d_a$ is not bounded, spans $(a,c)$ for at most $|C|=L-1$ values $c\in C$. Summing over the $k$ paths of $\cP$, there are at most $5k(L-1)$ pairs $(e,c)$ such that $c\in C$ and $e$ is a solution arc with an endpoint in $B$ that spans $(a,c)$. Since $|C|=L-1$, some $c\in C$ appears in at most $5k$ of these pairs.
\end{proof}

\section{The dynamic program}
\label{sec:dp}

A \emph{state} is a pair $(B,\sigma)$, where $B$ is a box and $\sigma$ is a set of at most $10k^2$ arcs crossing $B$, each with a \emph{color} $i\in[k]$ such that it is an arc of $D_i$. A \emph{piece} of color $i$ is a path of $D_i$ whose internal vertices are in $B$, whose first vertex is $s_i$ or not in $B$, and whose last vertex is $t_i$ or not in $B$. The state $(B,\sigma)$ is \emph{valid} if there are pieces with pairwise disjoint sets of vertices in $B$ such that, for every $i\in[k]$, the pieces of color $i$ contain the terminals $s_i$ and $t_i$ that are in $B$, and their arcs crossing $B$ are exactly the arcs of $\sigma$ of color $i$. In other words, the arcs of $\sigma$ and the terminals in $B$ can be linked by disjoint paths inside $B$, respecting the colors. As no arc crosses $B_0$, the state $(B_0,\emptyset)$ is valid if and only if there is a solution.

For every state $(B,\sigma)$, the algorithm computes a boolean value $T(B,\sigma)$, by increasing $|B|$:
\begin{itemize}
    \item if $|B|=1$, then $T(B,\sigma)$ is true if and only if $(B,\sigma)$ is valid. By Observation~\ref{obs:step}, no arc of $D_1\cup\dots\cup D_k$ has both endpoints in $B$, so every piece has at most one vertex in $B$, and this can be checked in polynomial time;
    \item otherwise, $T(B,\sigma)$ is true if and only if there are a cut of $B$ into $B^-$ and $B^+$ and a set $\tau$ of colored arcs, each with one endpoint in $B^-$ and the other in $B^+$, such that for each $\star\in\{-,+\}$, the set $\sigma^\star$ formed by $\tau$ and the arcs of $\sigma$ with an endpoint in $B^\star$ has at most $10k^2$ arcs and $T(B^\star,\sigma^\star)$ is true.
\end{itemize}
Note that $(B^\star,\sigma^\star)$ is a state. The algorithm answers yes if and only if $T(B_0,\emptyset)$ is true.

\begin{proof}[Proof of Theorem~\ref{thm:main}]
The distances $d_i$ and the digraphs $D_i$ can be computed in polynomial time by breadth-first search. Since the terminals are distinct, $k\le n$. There are $n^{\cO(k)}$ boxes, $n^{\cO(k^2)}$ sets $\sigma$ per box, and $n^{\cO(k^2)}$ choices of a cut and of $\tau$, so the running time is $n^{\cO(k^2)}$.

\emph{Soundness.} We show by induction on $|B|$ that if $T(B,\sigma)$ is true, then $(B,\sigma)$ is valid. If $|B|=1$, this holds by definition. Otherwise, consider the pieces given by induction for $(B^-,\sigma^-)$ and $(B^+,\sigma^+)$. Every arc of $\tau$ of color $i$ is the last arc of a piece of color $i$ on one side, and the first arc of a piece of color $i$ on the other side. Gluing the pieces along the arcs of $\tau$ gives walks of $D_i$, which are paths since $D_i$ is acyclic, and these are pieces showing that $(B,\sigma)$ is valid. In particular, if the algorithm answers yes, then $(B_0,\emptyset)$ is valid and there is a solution.

\emph{Completeness.} Let $\cP$ be a solution. For a box $B$, let $\sigma_\cP(B)$ be the set of solution arcs crossing $B$, each colored by the index of the path containing it. If $B$ is good for $\cP$, then $(B,\sigma_\cP(B))$ is a state, and it is valid, as shown by the maximal subpaths of $P_1,\dots,P_k$ whose internal vertices are in $B$. We show by induction on $|B|$ that $T(B,\sigma_\cP(B))$ is true for every box $B$ that is good for $\cP$. If $|B|=1$, this follows from validity. Otherwise, cut $B$ as in Lemma~\ref{lem:box_cut}, and let $\tau$ be the set of colored solution arcs between $B^-$ and $B^+$. Then $\sigma^\pm=\sigma_\cP(B^\pm)$ and $B^\pm$ is good for $\cP$, so $T(B^\pm,\sigma^\pm)$ is true by induction. Since $B_0$ is good for $\cP$ and $\sigma_\cP(B_0)=\emptyset$, the algorithm answers yes.
\end{proof}

\section{Improving the exponent}
\label{sec:klogk}

In this section, we sketch how to improve the running time to $n^{\cO(k\log k)}$.

\begin{theorem}\label{thm:klogk}
$k$-DDSPP can be solved in time $n^{\cO(k\log k)}$ on $n$-vertex digraphs.
\end{theorem}

The exponent $\cO(k^2)$ comes from the definition of good boxes, where the number of solution arcs crossing each of the $2k$ faces is bounded separately. When cutting $B$ into $B^-$ and $B^+$, only $\cO(k)$ crossing arcs are forced on the new faces around the cut, all the other crossings are inherited from $B$. If we could somehow guarantee that each new box inherits at most half of the arcs crossing its parent, it would be enough to create few new crossing arcs at each step and obtain a $k\log k$ bound. Let $H_k=\sum_{a=1}^k 1/a$ and $M=\lceil CkH_k\rceil$ for a large enough constant $C$. If a box is crossed by $w\le 4M$ solution arcs, and each new box is crossed by at most $w/2$ of them and at most $2M$ other solution arcs, then each new box is again crossed by at most $4M$ solution arcs.

To inherit at most half of the arcs, let $c$ be the coordinatewise median of the endpoints in $B$ of the arcs crossing $B$. For every $a\in[k]$, each of the open halfspaces $x_a<c_a$ and $x_a>c_a$ contains at most half of these endpoints, so a box lying in one of them inherits at most half of the arcs crossing $B$. We cannot simply cut $B$ at $c_a$, since the position of the cut is then fixed and many solution arcs may cross it. Moreover, the part containing $c$ lies on neither side of $c$, so it may inherit all the arcs crossing $B$. Instead, we surround $c$ by cubes centered at $c$, with radii roughly doubling from $1/4$ to more than $n$, and we partition each layer between two consecutive cubes into at most $2k$ boxes, each lying strictly on one side of $c$ in some coordinate (see Figure~\ref{fig:center}). Lemma~\ref{lem:layer} below shows that the radii and the partitions can be chosen so that, at every scale, at most $M$ solution arcs cross the cube and at most $M$ join two parts of the layer. The boxes of the layers are then crossed by at most $w/2+2M\le 4M$ solution arcs, and we recurse on them.

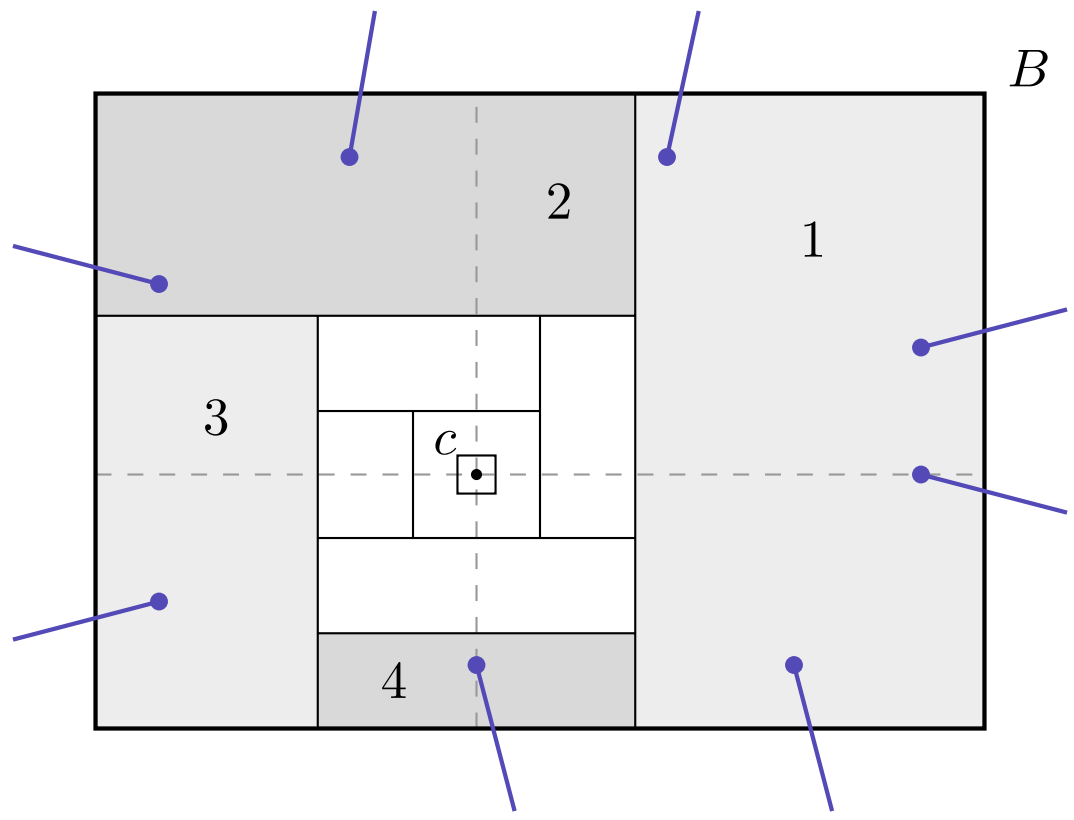
\begin{figure}[htb]
\centering
\begin{tikzpicture}[scale=0.42]
  \fill[black!7] (8.5,0) rectangle (14,10) (0,0) rectangle (3.5,6.5);
  \fill[black!15] (0,6.5) rectangle (8.5,10) (3.5,0) rectangle (8.5,1.5);
  \node at (11.3,7.7) {$1$};
  \node at (7.3,8.3) {$2$};
  \node at (1.9,4.9) {$3$};
  \node at (4.7,0.75) {$4$};
  \draw[dashed, black!40] (6,0) -- (6,10) (0,4) -- (14,4);
  \draw (8.5,0) -- (8.5,10) (0,6.5) -- (8.5,6.5) (3.5,0) -- (3.5,6.5) (3.5,1.5) -- (8.5,1.5);
  \draw (3.5,3) -- (8.5,3) (7,3) -- (7,6.5) (3.5,5) -- (7,5) (5,3) -- (5,5);
  \draw (5.7,3.7) rectangle (6.3,4.3);
  \draw[thick] (0,0) rectangle (14,10);
  \node at (14.7,10.4) {$B$};
  \fill (6,4) circle (2.5pt);
  \node at (5.5,4.5) {$c$};
  \foreach \ox/\oy/\mx/\my in {-1.3/7.6/1/7, -1.3/1.4/1/2, 4.4/11.3/4/9, 6.6/-1.3/6/1, 9.5/11.3/9/9, 11.6/-1.3/11/1, 15.3/6.6/13/6, 15.3/3.4/13/4} {
    \draw[pone, thick] (\ox,\oy) -- (\mx,\my);
    \fill[pone] (\mx,\my) circle (4pt);
  }
\end{tikzpicture}
\caption{The decomposition of a box $B$ for $k=2$. The dots are the endpoints in $B$ of the arcs crossing $B$, and $c$ is their coordinatewise median (dashed lines). The layers between nested squares centered at $c$ are partitioned into boxes, each lying strictly on one side of a dashed line, and thus containing at most half of the dots.}
\label{fig:center}
\end{figure}

\paragraph{Partitioning a layer.} Fix $c\in\mathbb{Z}^k$, and for $x\in\mathbb{R}^k$ and $a\in[k]$, let $y_a=x_a-c_a$. For $r>0$, let $Q(r)$ be the cube of points with $|y_a|\le r$ for every $a\in[k]$. Like a box, it has an upper and a lower $a$-face for every $a\in[k]$: a point is \emph{beyond} the upper (resp.\ lower) $a$-face if $y_a>r$ (resp.\ $y_a<-r$), and a point is outside $Q(r)$ if and only if it is beyond at least one face. To partition the points outside $Q(r)$, we fix an order $\pi$ on the $2k$ faces and take them one at a time: the first face in $\pi$ takes all the points beyond it, the second face takes the remaining points beyond it, and so on. In other words, we cut along the hyperplanes containing the faces of $Q(r)$, in the order $\pi$, each cut removing the part beyond one face. The parts are disjoint boxes covering the points outside $Q(r)$; in Figure~\ref{fig:center}, the numbers give the order for the outer layer. The \emph{label} of a point is the face that takes it, and $0$ if the point is in $Q(r)$. The points of a box with a given nonzero label form a box, as they are defined by one condition of the form $y_a>r$ or $y_a<-r$, and conditions of the form $y_b\le r$ or $y_b\ge -r$. Moreover, if the label is the upper (resp.\ lower) $a$-face, they all satisfy $x_a>c_a$ (resp.\ $x_a<c_a$), so they lie strictly on one side of $c$.

\begin{lemma}\label{lem:layer}
For every $c\in\mathbb{Z}^k$, $R>0$ and solution $\cP$, there are $r\in(R,2R)$ and an order $\pi$ such that at most $M$ solution arcs have exactly one endpoint in $Q(r)$, and at most $M$ solution arcs have both endpoints in $Q(4R)$ and endpoints with different labels.
\end{lemma}

\begin{proof}[Proof sketch]
Choose $r\in(R,2R)$ and $\pi$ uniformly at random and independently. Every arc counted in the statement has endpoints with different labels, and lies on the subpath of a path $P_i$ between its first and last vertices in $Q(4R)$, or is one of the two arcs just before and after it. We replace each arc $uv$ of $P_i$ by the two moves $u\to\min(u,v)\to v$, where the minimum is taken coordinatewise. On the first move all coordinates decrease, so the point can stop being beyond upper faces but cannot become beyond new ones. On the second move all coordinates increase, by at most one by Observation~\ref{obs:step}, so the point can become beyond upper faces but cannot stop being beyond them. Lower faces behave in the opposite way. If $u$ and $v$ have different labels, the label changes on one of the two moves.

For $a\in[k]$, let $C_a$ be the corner of points beyond at least $a$ upper faces, so that $C_1\supseteq\dots\supseteq C_k$. For instance, for $k=2$, $C_1$ is the set of points beyond the right or top face, and $C_2$ is the top right corner. We claim that leaving or entering $C_a$ costs at most $1/a$. Consider a move from $x$ to $x'$ on which all coordinates decrease, the other case being symmetric, and suppose that $x$ is beyond $s'$ upper faces and $x'$ beyond $s$ of them, so that the move leaves the corners $C_{s+1},\dots,C_{s'}$. The label of $x$ is the first in $\pi$ of the faces beyond $x$, so the move changes the label because of the upper faces only if this face is one of the $s'-s$ faces that are no longer beyond, which happens with probability $(s'-s)/s'\le 1/(s+1)+\dots+1/s'$. In other words, a point deep in a corner rarely changes label when it stops being beyond one of its faces. Lower faces are handled in the same way, with the corners of points beyond at least $a$ lower faces.

It remains to bound the number of times a path enters or leaves $C_a$. A point is in $C_a$ if and only if the $a$-th largest of its coordinates $y_1,\dots,y_k$ is larger than $r$. This value decreases on the first moves and increases by at most one on the second ones, since all coordinates do. Hence, as in the proof of Lemma~\ref{lem:movement}, it moves by $\cO(R)$ along the subpath, which has at most $8R$ arcs since $y_i$ increases by one along each arc, and the path enters or leaves $C_a$ only $\cO(1)$ times in expectation over $r$. Therefore, each path has $\cO(\sum_{a=1}^k 1/a)=\cO(H_k)$ arcs counted in the statement in expectation, and summing over the paths, some choice of $r$ and $\pi$ gives $\cO(kH_k)\le M$ such arcs. With a fixed order, entering or leaving a corner could cost $1$ each time, and we would only get $\cO(k)$ arcs per path.
\end{proof}

\paragraph{The decomposition.} Let $\cP$ be a solution and $B$ a box crossed by $w\le 4M$ solution arcs. Let $c$ be the coordinatewise lower median of the endpoints in $B$ of these arcs, or the image of any vertex of $B$ if $w=0$. Each coordinate $c_a$ is equal to $d_a(v)$ for some vertex $v$ in $B$. For $0\le h\le L=\lceil\log_2 n\rceil+3$, let $R_h=2^{h-2}$, let $(r_h,\pi_h)$ be given by Lemma~\ref{lem:layer} for $R=R_h$, and let $U_h=B\cap Q(r_h)$. Then $U_0$ contains at most one image of a vertex, and $U_L=B$. Since $r_h<r_{h+1}<4R_h$, the nonzero labels of $(r_h,\pi_h)$ partition $U_{h+1}\setminus U_h$ into at most $2k$ boxes, and at most $M$ solution arcs join two different parts of $U_{h+1}$. A solution arc crossing $U_h$ crosses $B$ or $Q(r_h)$, so at most $w+M\le 5M$ solution arcs cross $U_h$. Let $C$ be one of the new boxes. A solution arc crossing $C$ either crosses $B$ and has its endpoint in $B$ in $C$, or crosses $Q(r_{h+1})$, or joins $C$ to another part of $U_{h+1}$. Since $C$ lies strictly on one side of $c$ in some coordinate, at most $w/2+2M\le 4M$ solution arcs cross $C$. Moreover, $C$ has fewer distinct images than $B$, since no vertex $v$ of $C$ satisfies $d_a(v)=c_a$ for this coordinate $a$.

\paragraph{The dynamic program.} The algorithm is the same as in Section~\ref{sec:dp}, with the following changes. States are pairs $(B,\sigma)$ with $|\sigma|\le 4M$, processed by increasing number of distinct images in $B$. If $B$ contains at most one image of a vertex, $T(B,\sigma)$ is true if and only if $(B,\sigma)$ is valid, which can be checked directly. Otherwise, the center $c$ of a state is computed from $\sigma$ as above. The value $T(B,\sigma)$ is computed using an auxiliary table $T_\sigma(h,r,\alpha)$, where $r\in(R_h,2R_h)$ and $\alpha$ is a set of at most $5M$ colored arcs crossing $B\cap Q(r)$ that agrees with $\sigma$ on the arcs leaving $B$. For $h=0$, $T_\sigma(0,r,\alpha)$ is true if and only if $(B\cap Q(r),\alpha)$ is valid. For $h\ge 1$, $T_\sigma(h,r,\alpha)$ is true if and only if there are $r'\in(R_{h-1},2R_{h-1})$, an order $\pi$, and a set of at most $M$ colored arcs between the parts of $B\cap Q(r)$ defined by $(r',\pi)$, such that the state they determine on $B\cap Q(r')$ has at most $5M$ arcs and is true in $T_\sigma(h-1,r',\cdot)$, and the states they determine on the new boxes have at most $4M$ arcs and are true in $T$. As above, the new boxes have fewer distinct images than $B$. Finally, $T(B,\sigma)=T_\sigma(L,r,\sigma)$ for any $r\in(R_L,2R_L)$. Only the position of $r$ among the values $|d_a(v)-c_a|$ matters, so we only need $\cO(kn)$ radii per scale, and there are $(2k)!\le n^{\cO(k)}$ orders since $k\le n$. Soundness follows by gluing pieces as in Section~\ref{sec:dp}, and completeness follows by induction on the number of distinct images, using the decomposition above. All tables have size $n^{\cO(M)}$, so the running time is $n^{\cO(k\log k)}$.

{\small\setlength{\bibsep}{2pt}
\bibliographystyle{plainnat}
\bibliography{refs}
}

\end{document}